\documentstyle[epsf]{l-aa}
\input{psfig}
\def\ha{He~{\sc{i}}\ $\lambda$ 5876}
\def\hb{He~{\sc{ii}}\ $\lambda$ 4686}
\def\cc{C~{\sc{iv}}\ $\lambda\lambda$ 5801-5812}

\begin{document}
\thesaurus{06 (08.02.3; 08.23.2; 08.06.3; 11.19.5)}
\title{A dedicated photometric system for the classification of Wolf-Rayet stars\thanks{Based on observations collected at the European Southern Observatory, La Silla, Chile.}}
\author{P.\ Royer \thanks{Aspirant au Fonds National de la Recherche Scientifique (Belgium)} \and J.-M.\ Vreux \and 
J.\ Manfroid \thanks{Directeur de Recherche au Fonds National de la Recherche Scientifique (Belgium)}} 
\offprints{P. Royer, proyer@ulg.ac.be}
\institute{Institut d'Astrophysique, Universit\'e de Li\`ege,
5, Avenue de Cointe, B-4000 Li\`ege, Belgium}
\date{Received date; accepted date}
\maketitle

\newcommand{\rt}{$r_{\mbox{\tiny He I}}$}
\newcommand{\rn}{$r_{\mbox{\tiny He II}}$}
\newcommand{\rd}{$r_{\mbox{\tiny C IV}}$}
\newcommand{\ct}{$cont_{\mbox{\tiny He I}}$}
\newcommand{\cn}{$cont_{\mbox{\tiny He II}}$}
\newcommand{\cd}{$cont_{\mbox{\tiny C IV}}$}
\newcommand{\ltt}{$l(${\small He}~{\sc i}$)$}
\newcommand{\lnn}{$l(${\small He}~{\sc ii}$)$}
\newcommand{\ldd}{$l(${\small C}~{\sc iv}$)$}
\newcommand{\nwr}{non\ -\ WR}

\begin{abstract}
We present here tests of a five-filter photometric system aimed at WR classification.  In addition to the well-known easy separation between the WN and WC spectral types, these tests indicate interesting potentialities in the discrimination of subgroups among the WN and the WC which look well related to the classical subtypes. The proposed combinations of filters (or derived ones) bear enough discriminating power to satisfy some evolutionary studies in crowded fields where spectroscopic follow-up is not possible.
\end{abstract}

%
%
\section{Introduction}

Study of Wolf-Rayet (WR) star populations are important not only for our knowledge of the nature of the WR stars themselves but also for the invaluable information they give about their host galaxies and about galactic and stellar evolution in general.  The technological advances of the recent years allow to search much deeper in the absorbed regions of our own Galaxy as well as in an increasing number of external galaxies (Massey 1996 and references therein, Testor et al. 1996).

With a few exceptions (Corso, 1975; Armandroff \& Massey, 1985), the various surveys conducted so far consist of two-step procedures. The first one is a search for WR candidates by comparison of images obtained through various filters (Schild \& Testor 1991, Shara \& Moffat 1982, Shara et al. 1991) or by slitless spectroscopy (Azzopardi \& Breysacher 1979ab, Morgan \& Good 1985, Westerlund et al. 1983, Testor et al. 1996). The second is a spectroscopic study to precise the types and subtypes of the candidates.  However, spatial crowding rules out spectroscopic analysis in the densest regions of clusters and remote galaxies.  Clearly, these important fields of research are not compatible with a spectroscopic follow-up.

Using narrow-band filters, Corso (1975) was the first to achieve a differentiation between WN and WC spectral types (Massey \& Conti, 1983a).  Armandroff \& Massey (1985) improved the choice of narrow filters and used them to detect WR stars in Local Group galaxies and to separate the WN stars from the WC stars.  One of their figures indicates that their choice of filters also allows some rough selectivity between WNL and WNE but no selectivity among the WC subtypes. 


In this paper we present the first results of a five-filter photometric system whose physical content is high enough to allow meaningful comparisons with population modeling in the sense that it allows not only a separation between WN and WC spectral types, but also a discrimination at the level of subtypes, mainly for the WN subtypes.


%
%
\section{Selection of the filters}
Vreux et al. (1990), have shown that the sole use of the equivalent widths of He~{\sc ii} $\lambda$ 10124 and He {\sc i} $\lambda$ 10830 allowed to divide the WN population into six groups which are linked to the ``classical'' WN subtypes and to a ``temperature scale'' defined by the models of Hamann \& Schmutz (1987). It was then suggested that a four-band photometric system (two filters for the above mentioned lines, one for a carbon line such as C~{\sc iii} $\lambda$ 9715, and one for a continuum value) could lead to a useful classification system of the WR stars, particularly suited for crowded fields and for regions with a relatively high interstellar absorption.

Unfortunately, at that time, it quickly appeared that our access to an instrumentation suited for the 1 $\mu$m region was very limited. Consequently, a ``similar system'' using filters operating in the classical visible region was designed. Three ``line filters'' were selected, centered on \hb, \ha\ and \cc\, respectively.

The choice of the first two filters is obvious if one refers to the paper of Conti et al. (1990) where it is shown that an excellent correlation exists between, respectively, He~{\sc ii} $\lambda$ 10124 and He~{\sc ii} $\lambda$ 4686, and between He~{\sc i} $\lambda$ 10830 and He~{\sc i} $\lambda$ 5876.

The C~{\sc iv} line has been preferred to the much stronger nearby C~{\sc iii} $\lambda$ 5696 line because, in the same reference, it is shown that the C~{\sc iii} $\lambda$ 5696 line exhibits a ``peculiar behaviour'' relative to spectral types.

Since the three ``line filters'' are spread over about 1200~\AA, it is necessary to have two ``continuum filters''.  As is well known, finding an access to the continuum in a WR spectrum is not an obvious task. Here we have chosen two filters centered on $\lambda$ 5057 \AA\ and $\lambda$ 6051 \AA\ respectively. The first one avoids the {\small [O~{\sc III}]} $\lambda\lambda$ 4959-5007 nebular lines which could have been a cause of contamination in our diagrams. An a posteriori justification of the latter is given by Koesterke \& Hamann (1995) who consider that this wavelength gives the first access to a relatively clean ``continuum window'' in the WC stars, other windows being located at longer wavelengths.

The main characteristics of the filters discussed here are given in Table~1.

\begin{table}[hbt]
\begin{center}
\begin{tabular}{c c c c}
\hline\\
\bf Filter & \bf FWHM (\AA) & \bf Center (\AA) & \bf Emission line\\
\hline\\
\rn       &       30       &  4684  &  \hb\\
$c_1$     &       53       &  5057  &\\
\rd       &       27       &  5806  & \cc\\
\rt       &       26       &  5881  & \ha\\
$c_2$ 	  &       28       &  6051  &\\
\hline
\end{tabular}
\caption{Main characteristics of the filters}
\end{center}
\end{table}




%
%

\section{The data sets}

The various color diagrams and tests presented here are based on two distinct data sets. The first one consists of observations performed between 1991 and 1993 (some results are reported in Vreux et al. 1996). The second data set consists of spectrophotometric data from Torres-Dodgen \& Massey (1988, hereafter TM) for the WR stars and from Jacoby, Hunter \& Christian (1984, hereafter JHC) for the other stars.

\subsection{Observations}

The observations were conducted at the ESO La Silla observatory
during three dedicated runs using the ESO standard one-channel photometer
attached to the 1~m telescope. The photometer was equipped with
a Quantacon RCA 31034 tube. Data were obtained over a total of
26 nights : 6 in March 91, 13 in February 92 and 7 in September
93. Altogether 45 WR stars and 121 \nwr\ stars were observed during these observing runs.

The observing strategy was planned so as to fully benefit from the {\sc ranbo2} reduction algorithm (Manfroid 1993). The 121 \nwr\ constant stars were observed frequently in order to determine the photometric parameters. This was specially important because of the occasional presence of fast, large-amplitude, extinction variations due to volcanic aerosols. Except for a zero-point adjustment, the WR photometric system is 
the instrumental system. Due to the narrow passbands and the
stability of the filters, no color corrections were needed between the various runs. The zero of the system is fixed by the A0V star HD104430 for which 
the value 6.157 is imposed in every band.

Eight of the WR and most of the \nwr\ stars were not observed in the $c_1$  filter which we had not yet decided to include in the system. They could thus not be used in the subsequent color diagrams (Sect.~4.2).

\subsection{Spectrophotometry}

The TM spectrophotometric data base is made of 171 spectra of Galactic and Magellanic Clouds WR stars. Some of these spectra could not be used because their wavelength range was not sufficient. Individual inspection also led to the rejection of some spectra due to a too high noise level or to obvious anomalies in the relevant wavelength regions. Finally, we were left with 117 Galactic and Magellanic WR spectra.

The simulations presented in Sect.~5.1 also required the use of {\it normal}, i.e. \nwr\ stars spectra. This has been done through the use of a {\it representative subsample} of the JHC catalog, made of 28 stars.

\subsection{The complete data set}

The final data set consists of 129 WR stars. There are 37 galactic WR stars for which photometric data are available in all the needed filters and 117 WR stars for which spectrophotometric data are available (71 galactic and 36 magellanic WR stars). In this sample, there are 25 WR stars for which both photometric and spectrophotometric data are available. Altogether, this sample of 129 WR stars can be considered as highly representative as it covers almost 50\% of the whole presently known WR population of our Galaxy and the LMC.

In the present paper, {\it ``WR''} followed by a number will refer to a galactic WR star (van der Hucht et al. 1981) while {\it ``Brey''} followed by a number will refer to an LMC WR star (Breysacher 1981).

The \nwr\ sample consists of 64 objects : 36 of the 121 stars mentioned above, ranging from A to K and from luminosity class V to luminosity class III for which photometric observations have been performed, and 28 stars ranging from O to M and from dwarfs to supergiants (including Of stars) for which spectrophotometric data have been used.


%
%

\section{Color Diagrams}

When available, the spectral types attributed to the WN stars in all subsequent figures come from Smith, Shara \& Moffat (1996) and those of the WC stars come from Koesterke \& Hamann (1995). Otherwise, for the galactic stars, they come from van der Hucht et al. (1981) with additions or revisions by Massey \& Conti (1983b) while for the LMC stars, they come from Breysacher (1981), with additions or revisions by Massey \& Conti (1983c).  The reason of these choices is to include the weak-strong line distinction in our classification scheme.

The total data set consists of the results of the synthetic photometry performed on the spectra discussed above and of the results of the observing runs. In order to keep consistancy between Sects.~4 and~5, synthetic data were prefered to the observed ones when both were available (even though the precision is higher on the photometric data).



\begin{figure*}[htb]
\centerline{\psfig{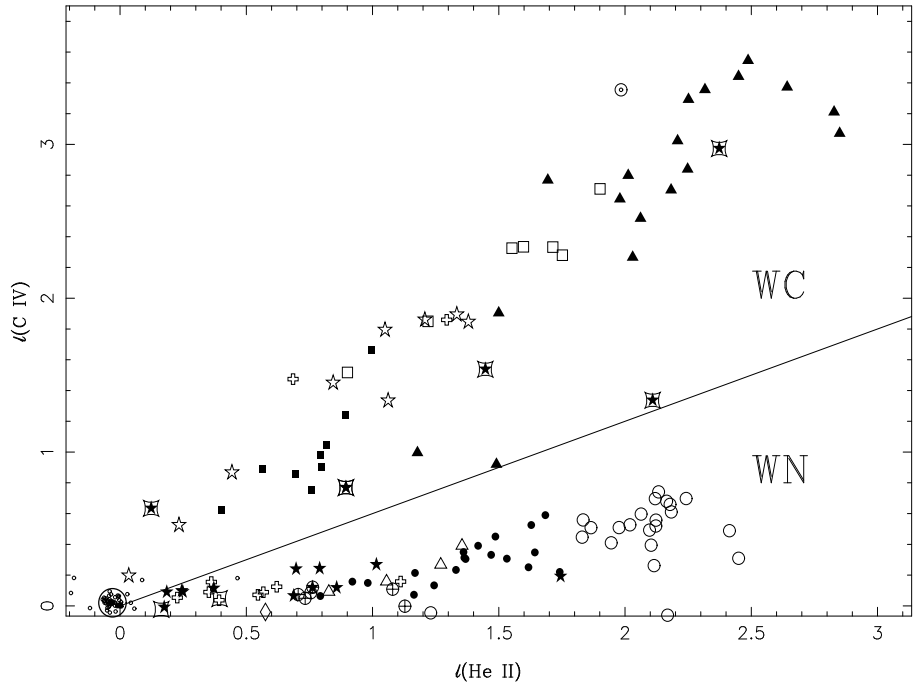}}
\caption{WN-WC separation. The symbols are explained in Table~3. The big circle around the origin encloses most of the \nwr\ stars.}
\end{figure*}
\begin{figure*}[htb]
\centerline{\psfig{figure=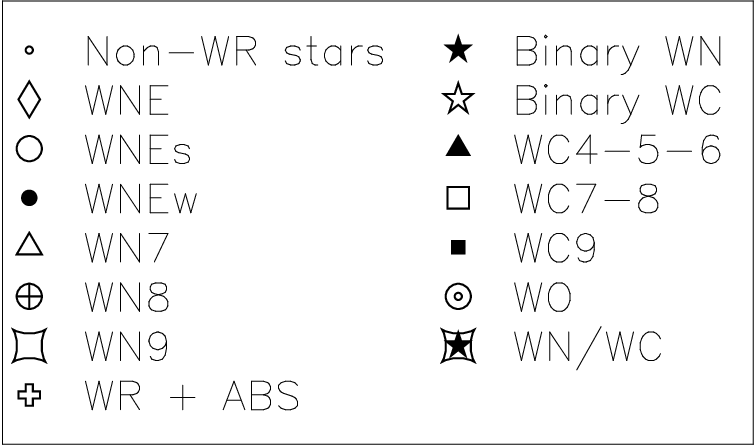,width=8cm,height=45mm,angle=270}}
\end{figure*}

\subsection{Calibration of the synthetic photometry}

In the whole data set, there are 25 WR stars with both observed and synthetic photometry. Moreover, two of the observed \nwr\ stars make part of the Hamuy et al. (1992) catalog of spectrophotometric standard stars (LTT~7379 and HR~7596)\footnote{These stars were not observed in the $c_1$ filter}. The calibration of the synthetic photometry was established on the basis of these 27 stars. A first iteration gave a mean zero point for each filter and deviations for every star. The most deviant stars (2~$\sigma$) were excluded and a second iteration provided the accepted zero points. Standard deviations are $\approx$ 0.1~mag in all filters which is compatible with the accuracy claimed in TM (Table~2).


\begin{table}[hbt]
\begin{center}
\begin{tabular}{c c c c}
\hline\\
\bf Filter & \bf s   & \bf n\\
\hline\\
\rn        &  0.021  &  24 \\
$c_1$      &  0.021  &  22 \\
\rd        &  0.019  &  24 \\
\rt        &  0.018  &  23 \\
$c_2$ 	   &  0.020  &  24 \\
\hline
\end{tabular}
\caption{Precision of the zero points of the synthetic photometry. $s$ is the standard deviation of the mean and $n$ the number of stars. We do not list the actual values which are unit dependent.}
\end{center}
\end{table}

\begin{table*}[htb]
\begin{center}
\begin{tabular}{c}
\end{tabular}
\vspace{-15mm}
{\bf Table.~3.} Symbols in the figures
\vspace{15mm}
\end{center}
\end{table*}
 
\subsection{The color indices}

Color diagrams based on very simple color indices, e.g., ($c_2$--\rd) vs ($c_1$--\rn), are able to separate WN stars from WC stars, but they do not discriminate them perfectly from \nwr\ stars, i.e. the constant stars observed to calibrate the photometric parameters of our system and the stars taken from JHC. To achieve a better separation, color indices are needed in which the continuum flux below the emission lines is more correctly taken into account.

Morris et al. (1993) indicate that, in the relevant wavelength domain, the continuum of WR stars can be approximated by a straight line in a $(\log{F_{\lambda}}$~vs~$\log{\lambda})$ diagram.  On the other hand, in the same wavelength domain, the interstellar reddening can be approximated by a straight line in a $(\log{F_{\lambda}}$~vs~$\lambda)$ diagram.  In both cases, the continuum under each line can be evaluated by a linear combination of the $c_1$ and $c_2$ magnitudes.  The difference between the two approximations appears only in the values of the parameters included in the linear combinations.  Nevertheless, simulations indicate that these differences are quite small because the wavelength domain concerned is rather limited ($\approx 1400$ \AA). In this paper, the continuum magnitudes under the different lines were computed in the framework of the second approximation here above $(\log{F_{\lambda}}$~vs~$\lambda)$, i.e. with the following relations :

\begin{description}
 \item[]   \cn\ $= c_1 + 0.373(c_1$--$c_2)$
 \item[]   \cd\ $= c_2 + 0.2475(c_1$--$c_2)$
 \item[]   \ct\ $= c_2 + 0.177(c_1$--$c_2)$
\end{description}

The indices linked to the normalized intensities of the lines are defined as :

\begin{description}
 \item[] \lnn\ = \cn\ - \rn
 \item[] \ldd\ = \cd\ - \rd
 \item[] \ltt\ = \ct\ - \rt
\end{description} 

\subsection{Separation between \nwr\ stars, WN and WC stars}

This separation is easily achieved by a plot of \ldd~vs~\lnn, as shown in Fig.~1.  The meaning of the various symbols is given in Table~3. Figure~1 shows a neat separation between the WC stars and WN stars, which are distributed along two distinct branches. Some kind of rough classification is operated along both branches and the WN/WC stars lie either in the WC branch or between both branches. It is interesting to note the unusually large range of the color indices compared to other more conventional systems, e.g., $UBV$.

In Fig.~1, as in every subsequent ones, the normal stars cluster near the origin because they have no strong features in the filter passbands. Incidentally, this implies that binaries containing a WR star and a normal star are expected to be closer to the origin than the single WR stars. The only \nwr\ stars that do not really belong to that cluster of points are K and M stars and supergiants Of. However, nearly all of them do appear in regions where they cannot be mismatched with WR stars (too close to the cluster of other \nwr\ stars or even not in the same quadrant as the WR stars). The only \nwr\ star of the simulations that seems hard to disentangle from the WN stars is a dwarf M1 (located at \lnn $\approx$ 0.5, \ldd\ $\approx$ 0.2). A giant K4 (at \lnn\ $\approx$ 0.1, \ldd\ $\approx$ 0.2) is also badly placed on Fig.~1 but we will see that it can be rejected on the basis of Fig.~2 because all \nwr\ stars (the M1V excepted) lie at \ltt~$<$~0.1 and \lnn~$<$~0.1. The measurable positive value of the \lnn\ index in the case of the M1V star finds its origin in a molecular absorption band around our $c_1$ filter which depletes the reference continuum used to evaluate the \hb\ emission.



\begin{figure*}[hb]
\centerline{\psfig{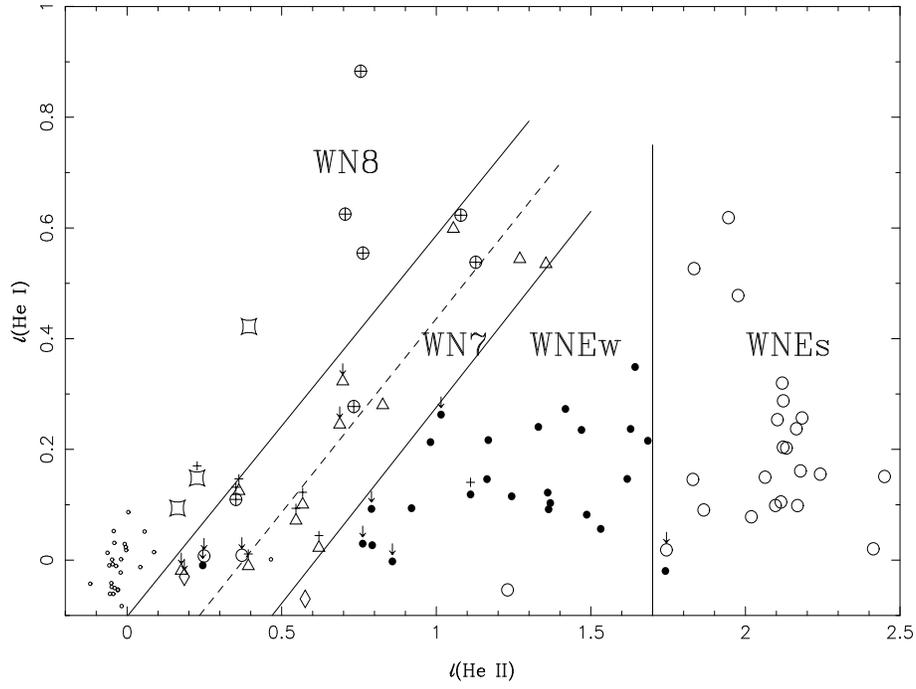}}
\caption{Intra-WN separation. A downward arrow indicates a binary. A plus sign above another symbol indicates a  ``+ abs'' star. The other symbols are explained in Table~3.}
\end{figure*}
\begin{figure*}[ht]
\centerline{\psfig{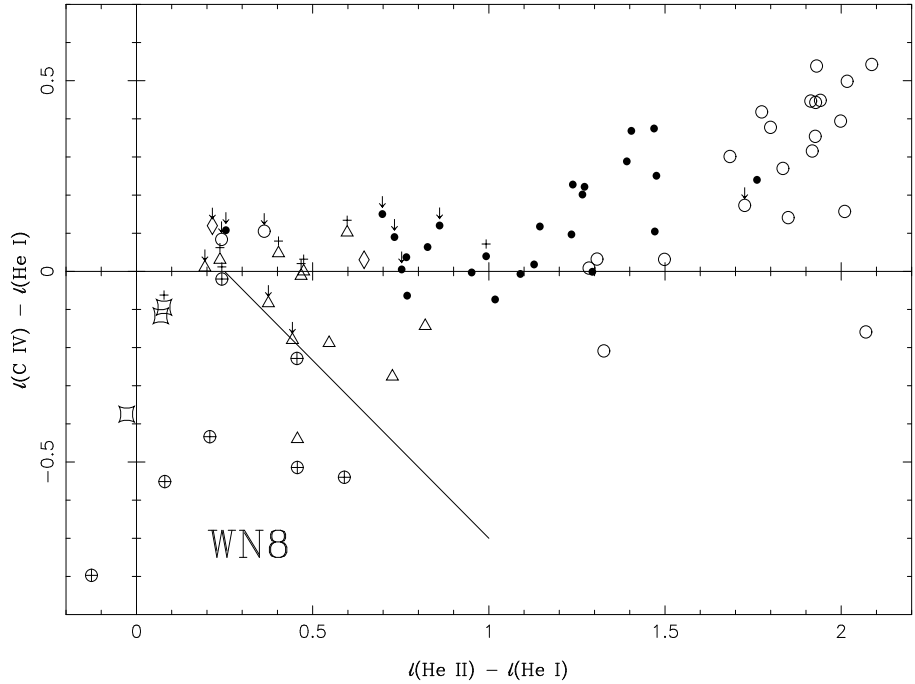}}
\caption{WN8-9 separation based on line ratios. A downward arrow indicates a binary. A plus sign above another symbol indicates a  ``+ abs'' star. The other symbols are explained in Table~3.}
\end{figure*}

\subsection{Intra-WN separations}

Figure~2 shows a plot of \ltt~vs~\lnn, with the sample limited to the WN stars as selected from Fig.~1 and the synthetic photometry on \nwr\ stars. In this figure, the binaries and the WN+abs are plotted with the symbol corresponding to the spectral type of their WR member, except that the former are identified by a downward arrow and the latter by a plus sign.  This graph shows a rather clear separation between the WNEs stars, the WNEw stars, and the WNL stars. Among the latter, there is also a tendency to have a segregation of the WN7 stars and the WN8-9 stars in two separate regions, with an intermediate zone. As expected, the binaries are clustering around the origin and the horizontal axis of the plot. Indeed, almost all of the stars nearest to the origin are known binaries or exhibit absorption lines in their spectra. The only exception is Brey 58 (diamond shaped point at \lnn\ $\approx$ 0.6). This star has a rather uncertain classification : WN5-6 or Of according respectively to Breysacher (1981) or Smith et al. (1996). Some WN9 stars also lie close to the \nwr\ star region which is normal as most of these objects are considered as closely related to the LBV stars, i.e., to be representative of a transition stage between Of and WN stars.


It may be interesting to mention that, when plotted on this diagram, the WN/WC stars place themselves according to their WN features, except WR 26 (WN7s/WCE, at \lnn\ $\approx$ 2.4 and \ltt\ $\approx$ 1.1) which exhibits strong WC characteristics (Fig.~1) and Brey~72 which is not a single object (Breysacher 1981). We get :
\begin{itemize}
 \item WR~98 (WN8w/WC7) in the WN8-9 stars
 \item WR~8 (WN7w/WCE) in the WN7 stars
 \item WR~58 (WN4s/WCE) in the strong-lined WNE stars
\end{itemize}

A better separation between the WN8 stars and the WN7 stars is achieved on the color diagram in Fig.~3 which uses more complex color indices. On this graph, we no longer have an intermediate strip where the WN8-9 stars and the WN7 stars are mixed. WR~12, recently reclassified as a WN8 (Rauw et al. 1996; Eenens et al. 1996; Smith et al. 1996), is lying well within the WN8-9 part of the plot. The only WN7 star lying in this same region is WR~82 that Smith et al. reclassify as a weak-lined WN7 star but which was previously known as a WN8 star.

On this diagram, all the WNEw stars lying at \lnn~-~\ltt~$<$~0.9 are binaries, except Brey~60, classified as peculiar WN3 (Smith et al. 1996) and WR 28 and Brey 47, which have recently been subject to a reclassification by Smith et al. (1996).


\begin{figure*}[htb]
\centerline{\psfig{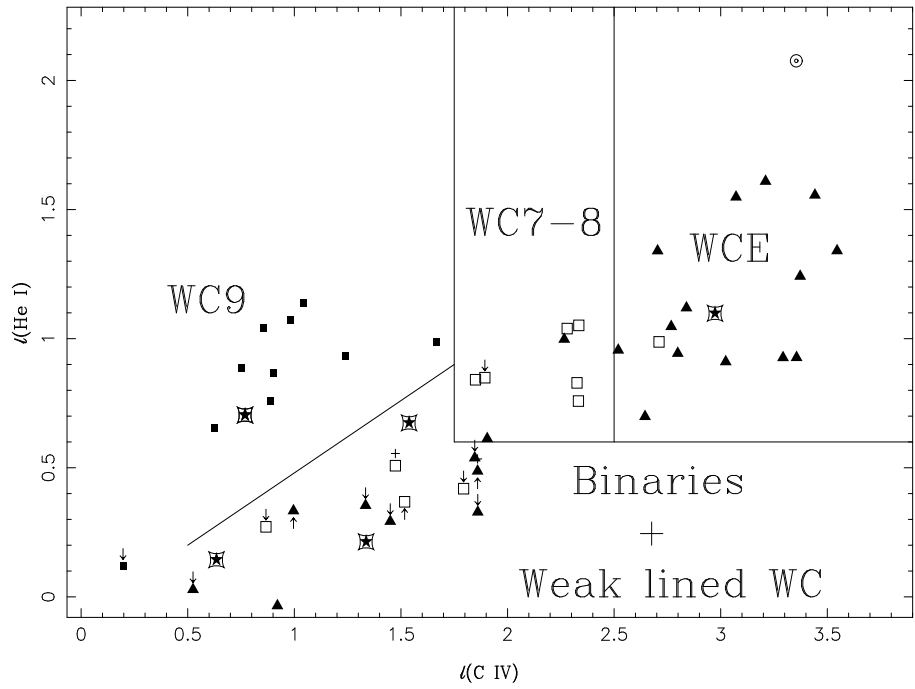}}
\caption{Intra-WC separation.  A downward arrow indicates a binary. An upward arrow indicates a weak lined WC star. A plus sign above another symbol indicates a  ``+ abs'' star. The other symbols are explained in Table~3.}
\end{figure*}

\subsection{Intra-WC separations}

In Fig.~4, the \ltt\ index is plotted as a function of \ldd. Some groups can easily be identified : the WC9 stars are located in the upper left corner while the WCE stars lie in the upper right corner, though not far apart from the WC7~-~8 stars. The representative point of the WO (Brey~93~= Sand~2) shows up near the WCE region, but clearly apart from it.

The lower part of the diagram is occupied by the binaries, the 3 weak-lined WC stars of our sample (WR~39~: WC6w; WR~86~: WC7w and WR~50~: WC6w+abs), plus WR~72, classified as peculiar WC4.

Concerning the WN/WC stars, the situation is reversed in comparison to Fig.~2 since only the strong-lined WR26 is located according to its WC features.


\section{Additional characteristics}

\begin{figure*}[htb]
\centerline{\psfig{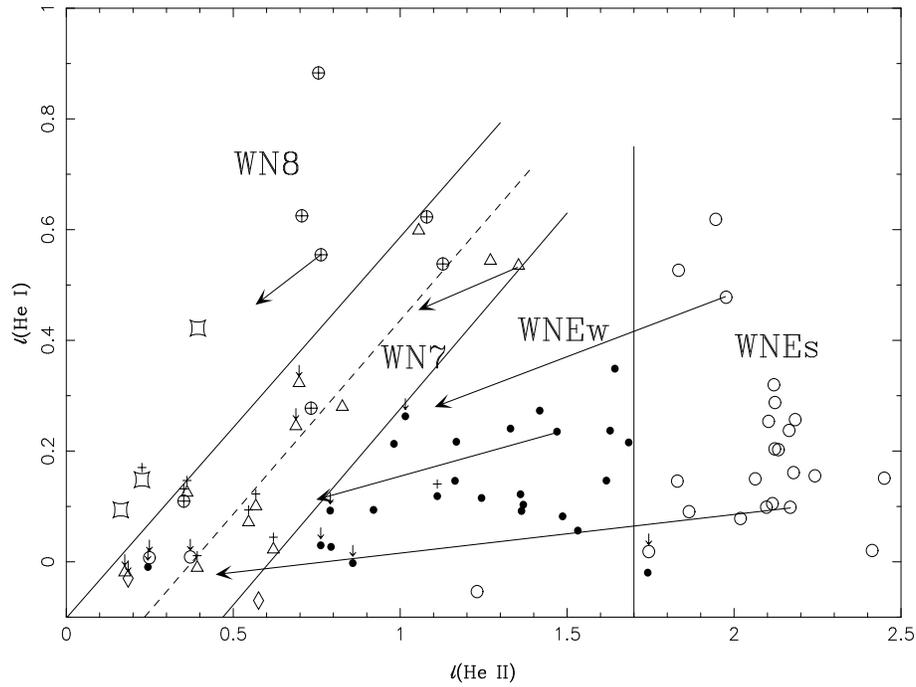}}
\caption{Simulations for WN binaries.  A downward arrow above another symbol indicates a binary. A plus sign above another symbol indicates a  ``+ abs'' star. The other symbols are explained in Table~3. The longer arrows indicate the displacement imposed to the representative points of some WN stars when their light is mixed with that of an O6V companion star (see text for details).}
\end{figure*}
\begin{figure*}[htb]
\centerline{\psfig{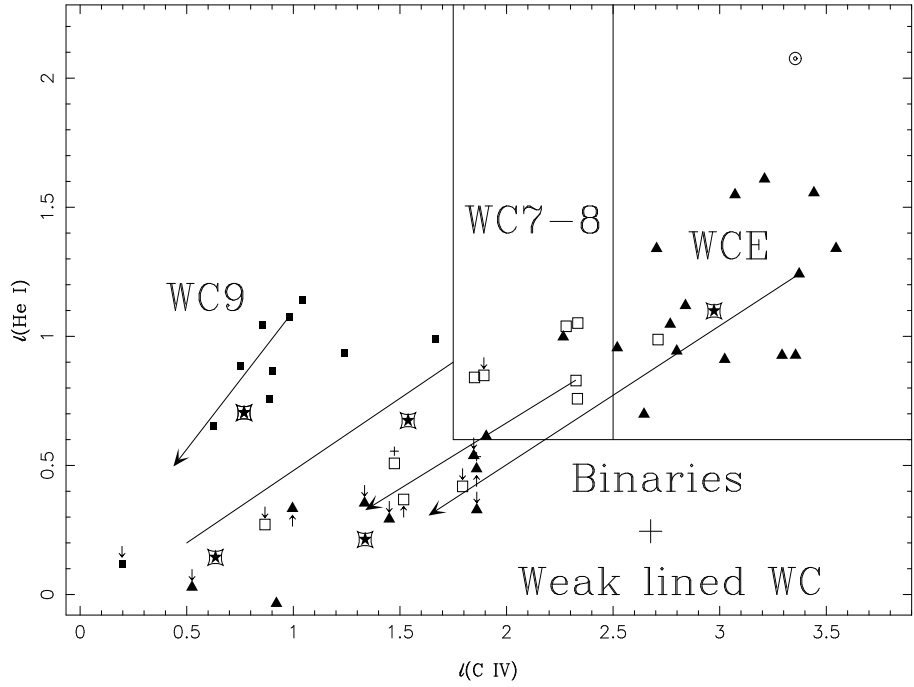}}
\caption{Simulations for WC binaries.  A downward arrow above another symbol indicates a binary. A plus sign above another symbol indicates a  ``+ abs'' star. An upward arrow below another symbol indicates a weak lined WC star. The other symbols are explained in Table~3. The longer arrows indicate the displacement imposed to the representative points of some WC stars when their light is mixed with that of an O6V companion star. From left to right, the WR stars used for the simulations are WR~103 (WC9), WR~57 (WC7) and Brey~8 (WC5-6).}
\end{figure*}

\subsection{Influence of a companion}
The observed WR stars may be physical (typically WR~+~OB) or spurious (WR~+~anything on the line of sight) binaries\footnote{For the sake of simplicity, the second star will be called {\it companion} in any case}. In both cases, the spectral lines of the WR stars are ``diluted'' by the other star's light, and the corresponding location of these stars on our color diagrams lie closer to the origin of the axis than the corresponding uncontaminated spectral types (Sect.~4.3). This is indeed observed on Figs.~1 to~4 where the symbols corresponding to the binaries lie closer to the origin than those representing single objects.

The first interesting fact to notice is that the binarity has no effect on the WN-WC separation as can be seen on Fig.~1. This means that, even when dealing with spatially unresolved double (or even multiple) objects, photometry alone still allows not only to detect the WR stars but also to separate them into WN and WC stars. But, in most cases, much more can still be achieved. As can be seen in Fig.~2, most of the binaries lie in their proper zone. It is true for the WN7 stars and it is also true for the WNEw stars with only one exception, WR 43, a WNEw star which is lying in the intermediate strip between the WN7 and the WN8-9 regions, with \lnn\ $\approx$ 0.25. As a matter of fact, this star has some reasons to be mislocated on our diagram since Hofman et al. (1995) have shown that, within a circle of 5'' diameter around the star, diffraction-limited speckle masking observations allow to count 20 different objects. Among them, 4 are of the WN type, the others probably of type Of. It is interesting to see that even in such a difficult situation, photometry still allows to classify the object as a WN star.

The other ``mislocated'' objects on Fig.~2 are Brey 40A (the diamond at \lnn\ $\approx$ 0.2), a WN3+O6 (Niemela 1991), Brey 21 (the open circle at \lnn\ $\approx$ 0.25), a WN5?+B1I (Smith et al. 1996) and WR 97 (the open circle at \lnn\ $\approx$ 0.38), a WN5b + 07 (Smith et al. 1996). Due to their low intrinsic luminosity ($M_{\mbox{\sc v}} \approx -3.$, van der Hucht 1992), it is expected that the contamination of a WN3 star by an O6 star ($M_{\mbox{\sc v}} \approx -5.5$) shifts markedly the representative point towards the origin. It is more surprising for a WN5 star with a luminosity of the order of $M_{\mbox{\sc v}} \approx -5.$ (van der Hucht 1992). To quantify this assertion, we have simulated the contamination of the WR stars of our sample by an O6 object. 

As can be seen on Fig.~5, the contamination of a strong lined WN3 star (Brey 1, located at \lnn\ $\approx$ 2.2) by a single O6 star brings its representative point close to the location of Brey 40A. But the contamination of a strong lined WN6 star (WR 110, located at \lnn\ $\approx$ 2) leads to a markedly shorter displacement which is not sufficient to bring a representative point from the strong lined WNE region to the vicinity of the origin of the graph. This could indicate either that the WR component of Brey 21 and WR 97 is less luminous than a classical strong lined WNE star or that the WR light is diluted by more than one companion.

The results of a few other simulations are also shown on the graph. As expected, due to their high luminosities, the representative points of the late WN stars are less affected by the presence of a companion. It is also interesting to notice that the nature of the displacements is such that, basically, a contamination maintains roughly the representative points within their spectral region : a WN8-9 star will always remain in its zone, a WN7 star will roughly do the same, and it is only close to the origin that an early WN star could enter the WN7~+~abs region. All the preceding conclusions are not affected if the simulations are made with another spectral type for the companion because the displacement of the representative point comes from the dilution of the WR spectral lines, not from the spectral characteristics of the companions.

Similar simulations have been carried out with WC objects (Fig.~6). Again, the displacement is towards the origin of the axes, i.e. such as to bring the representative points of the binaries in the binaries region, as defined on Figs.~4 and 6. The length of this displacement is a function of the relative luminosities of the WR star and its companion star. It is interesting to notice that the WC9 stars are so well separated from the other WC stars that most of their binaries probably remain in a distinct region of the plot.

\subsection{Influence of reddening}
An important fact is that these results are little affected by reddening (simulated up to $A_{\mbox{\sc v}} = 7.$). Its effect is indeed weak and roughly the same for all WR and normal stars, so that differential reddening in a given part of the sky could blur a bit the separations between the different regions in our color diagrams, but not in a drastic way.

\subsection{Radial velocity shifts}
Observing external galaxies with narrow band filters raises the question of adapting the filters to the radial velocities of the targets. In order to estimate to which extent our results were sensitive to these velocities, we performed synthetic photometry with nominal and enlarged \lnn, \ldd\ and \ltt\ filters (by a factor of 2) on blue- and red- shifted WR spectra.

Results are encouraging since, thanks to the width of the WR lines, blueshifts are no source of major problems and redshifts can be tolerated up to several hundreds of km/s. Of course, the useful part of our plots slightly shrinks as the redshift increases, i.e. the lines begin to get out of the filters and the stars slowly travel to the origin of the diagram, but sufficient discriminating power is left to operate till $\sim$ 500 km/s redshifts. The first property to suffer from further redshift is the WC9-WN separation. Apart from that, only slight adaptation of the borders of the different regions could be necessary in the \ldd~vs~\ltt\ plot at such redshifts.

Tests performed with wider band pass bring no surprise as they prove these to be less sensitive to redshift, but they also reduce discriminating power because the light of the line is more and more diluted by the continuum light as the filters widen.

\section{Conclusions}

As previous WR oriented photometric systems, the combination of filters presented here allows a good discrimination between WR stars and stars of other spectral types, as well as an easy distinction between the WN and the WC stars. In addition to this primary classification, it has been shown that the system proposed here has a much deeper discriminating power. This is particularly true for the WN subtypes for which a separation between WN8, WN7, WNEw and WNEs is achieved. Even if less discriminating power exists for the WC stars, WC9 stars, single WC7-8 stars and WCE stars are easily identified.

An examination of the plots as well as various simulations show that not only the WN/WC distinction but also the separation between the WN subtypes is basically maintained even when the light of the WR star is diluted by that of a possible companion, the only exception being the low-luminosity early subtypes.

The separation between the WN8 stars and the WN7 stars as shown on Fig.~3 is particularly interesting since recent works on the filiations between the WR subtypes indicate that such a discrimination is physically important. Indeed, the WN7 stars are presently considered to be the descendants of objects more massive than $60 M_{\odot}$ while the WN8 stars are descendants of $40$ to $60 M_{\odot}$ progenitors (Maeder 1996). Consequently, discriminating these two subclasses can bring information on the mass distribution of the progenitors, i.e. on the initial mass function (IMF) of the observed region.

Combined with state of the art techniques to deconvolve crowded fields, the photometric system discussed here should allow to get information on the upper part of the IMF in the cores of H{\sc ii} regions, well beyond the limit reachable through the usual spectroscopic methods.

\acknowledgements{The authors are indebted to the Fonds National
de la Recherche Scientifique (Belgium) for multiple supports.
This research is also supported in part by contract ARC\,94/99-178 ``Action de recherche concert\'ee de la Communaut\'e Fran\c{c}aise'' (Belgium). Partial support through the PRODEX XMM-OM Project is also gratefully acknowledged. We thank our referee Jacques Breysacher for a careful reading of the manuscript.}

\end{document}